\documentclass[twocolumn,superscriptaddress,colorlinks=true,allcolors=blue]{revtex4-2}
\pdfoutput=1

\usepackage{silence}
\usepackage{comment}

\usepackage[utf8]{inputenc}
\usepackage[breaklinks]{hyperref}
\usepackage[ 
    paperwidth=210mm,
    paperheight=276mm,
    textwidth=180mm,
    textheight=243.5mm,
    columnsep=4mm, 
]{geometry}
\usepackage[british]{babel}
\usepackage[babel=true]{csquotes}

\usepackage{graphicx}
\graphicspath{{figures/}}
\usepackage{xcolor}

\usepackage{amsmath}
\usepackage{siunitx}
\DeclareSIUnit{\fps}{ \translate{frames per second} }
\usepackage{textgreek}
\usepackage{textcomp}
\usepackage{lmodern}
\usepackage{tabularx}
 
\setcitestyle{super} 

\usepackage{titlesec}
\titleformat*{\section}{\raggedright\bfseries\sffamily\large}
\titlespacing*{\section}{0em}{1em}{0.5em}
\titleformat{\subsection}[runin]{\raggedright\bfseries}{}{}{}[.]
\titlespacing*{\subsection}{0em}{1em}{0.5em}
\usepackage{lettrine}

\usepackage{caption}
\newcommand{\tas}{\texorpdfstring{1\textit{T}-TaS\textsubscript{2}}{1\textit{T}-TaS2}}
\usepackage{float}

\begin{document}

\title{\sffamily\Large\texorpdfstring
{Light-induced hexatic state in a layered quantum material}
{Light-induced hexatic state in a layered quantum material}
}

\author{Till Domröse}
\author{Thomas Danz}
\affiliation{Max Planck Institute for Multidisciplinary Sciences, 37077 Göttingen, Germany}
\author{Sophie F. Schaible}
\affiliation{4th Physical Institute -- Solids and Nanostructures, University of Göttingen, 37077 Göttingen, Germany}
\author{Kai Rossnagel}
\affiliation{Institute of Experimental and Applied Physics, Kiel University, 24098 Kiel, Germany}
\affiliation{Ruprecht Haensel Laboratory, Deutsches Elektronen-Synchrotron DESY, 22607 Hamburg, Germany}
\author{Sergey V. Yalunin}
\affiliation{Max Planck Institute for Multidisciplinary Sciences, 37077 Göttingen, Germany}
\author{Claus Ropers}
\email[Corresponding author: ]{claus.ropers@mpinat.mpg.de}
\affiliation{Max Planck Institute for Multidisciplinary Sciences, 37077 Göttingen, Germany}
\affiliation{4th Physical Institute -- Solids and Nanostructures, University of Göttingen, 37077 Göttingen, Germany}


\begin{abstract}
\noindent  The tunability of materials properties by light promises a wealth of future applications in energy conversion and information technology. Strongly correlated materials such as transition-metal dichalcogenides (TMDCs) offer optical control of electronic phases, charge ordering and interlayer correlations by photodoping. Here, we find the emergence of a transient hexatic state in a TMDC thin-film during the laser-induced transformation between two charge-density wave (CDW) phases. Introducing tilt-series ultrafast nanobeam electron diffraction, we reconstruct CDW rocking curves at high momentum resolution. An intermittent suppression of three-dimensional structural correlations promotes a loss of in-plane translational order characteristic of a hexatic intermediate. Our results demonstrate the merit of tomographic ultrafast structural probing in tracing coupled order parameters, heralding universal nanoscale access to laser-induced dimensionality control in functional heterostructures and devices.
\end{abstract}

\maketitle

\lettrine[lines=3,nindent=0.5em,findent=-0em]{C}{} ollective behaviour beyond the single-particle picture fosters a variety of fundamental physical phenomena, such as superconductivity \cite{Orenstein2000} and ferromagnetism \cite{Spaldin2005}, which involve the establishment of long-range order below a critical temperature. Due to weaker screening and reduced phase space volumes for scattering, collective excitations are particularly important in low-dimensional systems. Prominently, plasmon and spin waves constitute elementary excitations of the one-dimensional Luttinger liquid \cite{Voit1995}.

For ideal systems of reduced dimensionality \cite{Li2021}, the \textit{Mermin-Wagner-Hohenberg} theorem prohibits long-range ordering at finite temperatures, owing to thermal fluctuations \cite{Hohenberg1967}. Strictly speaking, this precludes symmetry-breaking phase transitions. Instead, the transition from a quasi-long-range ordered two-dimensional solid to a liquid phase is described within the framework of \emph{Kosterlitz-Thouless-Halperin-Nelson-Young} (KTHNY) theory \cite{Kosterlitz2016,Nelson1979}. While the solid-liquid transition is always of first order in three-dimensional systems \cite{Birgeneau1986}, the theory predicts two successive second-order transitions in two dimensions via a hexatic phase. This intermediate is characterised by intact orientational but reduced translational order arising from the presence of unbound topological defects \cite{Nelson1979,Gasser2010}.

Experimental realisations of this transition have been actively pursued with two-dimensional model systems of colloidal spheres \cite{Gasser2010}, particles physisorbed on crystalline substrates \cite{Birgeneau1986}, skyrmion lattices \cite{Huang2020} and technologically relevant smectic liquid crystals \cite{Zaluzhnyy2022}. Yet, quasi-two-dimensional character is also readily found in layered materials \cite{Li2021}. Among those, transition metal dichalcogenides (TMDCs) stand out due to their highly anisotropic correlations \cite{Manzeli2017}. This leads to the occurrence of direct-to-indirect band gap transitions \cite{Wang2012b}, layer- and doping-dependent superconductivity \cite{Sipos2008,Xi2015,Luo2016,Ugeda2016}, and the formation of charge-density waves (CDWs) \cite{Rossnagel2011}. Due to the delicate interplay between the various degrees of freedom, the properties of these systems, often referred to as \textit{quantum materials} \cite{Tokura2017}, are highly tunable by external stimuli \cite{Fausti2011,Poellmann2015,Borzda2015,Sie2019,Bao2022,Schmitt2021}.

\begin{figure}[b]
\centering
\includegraphics[width=\columnwidth]{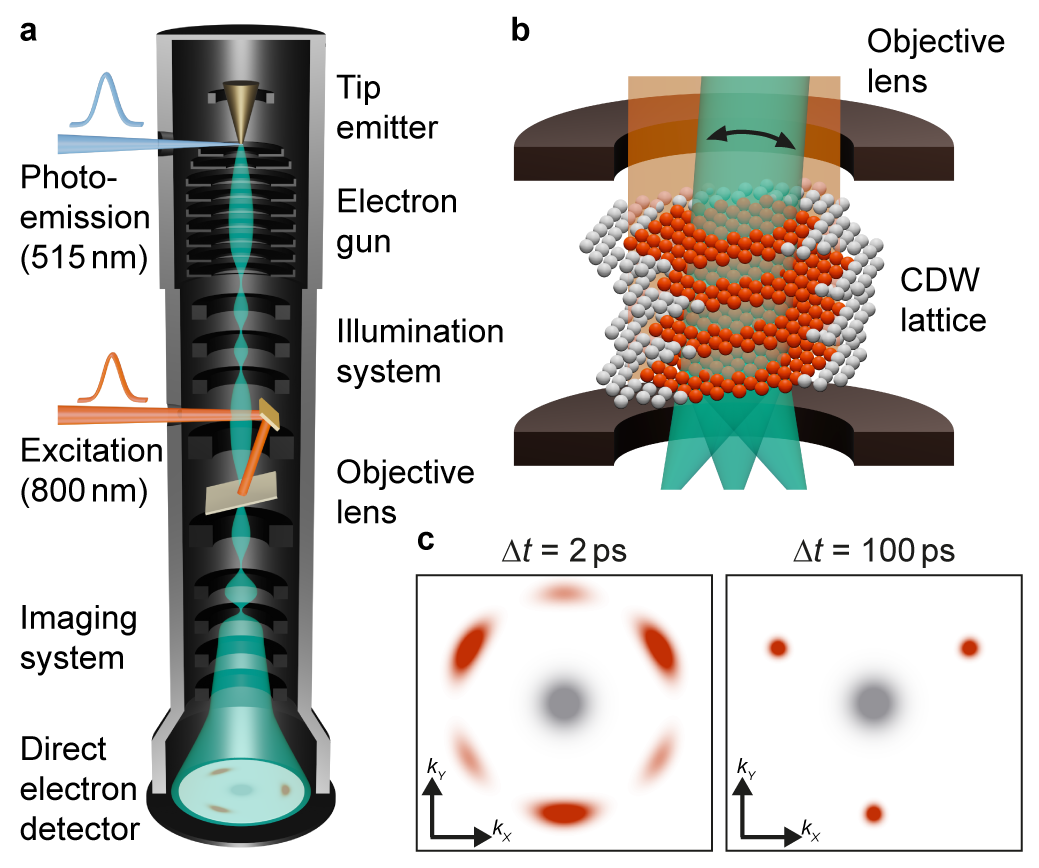}
\caption{
\textbf{High-coherence ultrafast nanobeam diffraction.}
\textbf{a}~Schematic of the Ultrafast Transmission Electron Microscope (UTEM). A specimen (inset in \textbf{b}) is excited using ultrashort laser pulses (red). After a variable temporal delay $\Delta t$, ultrashort electron pulses (green) are emitted from a tip-shaped photocathode via linear photoemission (blue) and capture the specimen's transient state on a direct electron detection camera (inset in \textbf{c}).
\textbf{b}~Visualisation of the \tas{} specimen located within the objective lens of the UTEM instrument. The optical excitation drives a transformation of the layered CDW lattice between two incommensurate CDW phases (represented by grey and red spheres), probed by a nanometer-sized electron beam of variable tilt (double arrow).
\textbf{c}~Schematic diffraction patterns for two temporal delays. CDW spots (red) are arranged around a main lattice peak (grey). The temporal evolution of CDW spot intensities and profiles (radial as well as azimuthal) encodes both in- and out-of-plane CDW dynamics on ultrashort timescales.
}
\label{fig:1}
\end{figure}

\begin{figure*}[t]
\centering
\includegraphics[width=\textwidth]{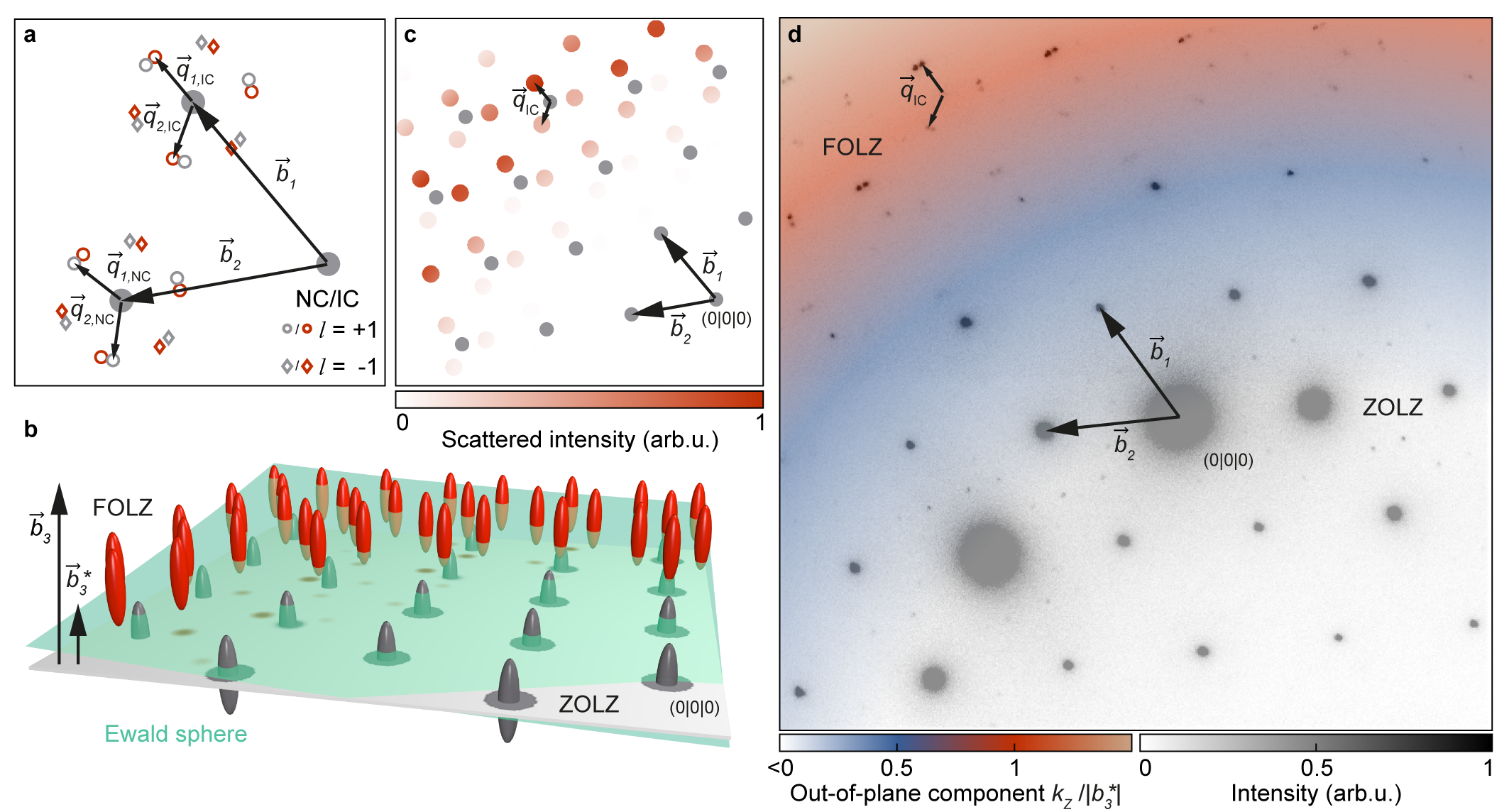}
\caption{
\textbf{Reciprocal lattice of \tas{}, diffraction geometry and structure factor.}
\textbf{a}~Reciprocal lattice of the NC (open grey) and IC phase (red) of \tas{} (first-order spots only). Due to the CDW stacking periodicity, the CDW spots are located in the first-order Laue zone (FOLZ, $l=3n\pm 1$ in Eq.~\ref{eq:Spijkerman}) of the CDW lattice, while the related host lattice reflections (solid grey) lie in the zero-order Laue zone (ZOLZ).
\textbf{b}~Ewald construction visualising the diffraction geometry. The intensity scattered into individual first-order CDW spots is governed by proximity to the Laue condition, i.e., where the Ewald sphere (green) intersects the CDW reciprocal lattice rods (red). Accessing the FOLZ requires an electron beam slightly tilted away from the $[001]$ zone axis. The low curvature of the Ewald sphere for \SI{120}{\keV} electron energy results in a dense sampling of the rod shape.
\textbf{c}~A second important contribution to the CDW spots intensity is given by the CDW structure factor depending on the angle between scattering vector and CDW wave vector (see Supplementary Information).
\textbf{d}~Temporally averaged experimental diffractogram. The high electron beam coherence leads to a clear separation of NC and IC spots. The coloured overlay indicates the local height $k_z$ of the Ewald sphere above the ZOLZ in units of $\vec b_3^*$.
While first-order spots of both phases are the dominant feature in the FOLZ and contain information on the stacking periodicity, the low-intensity second-order NC spots appearing between bright host reflections in the ZOLZ are a sensitive measure of the CDW amplitude.
}
\label{fig:2}
\end{figure*}

\begin{figure*}[t]
\centering
\includegraphics[width=\textwidth]{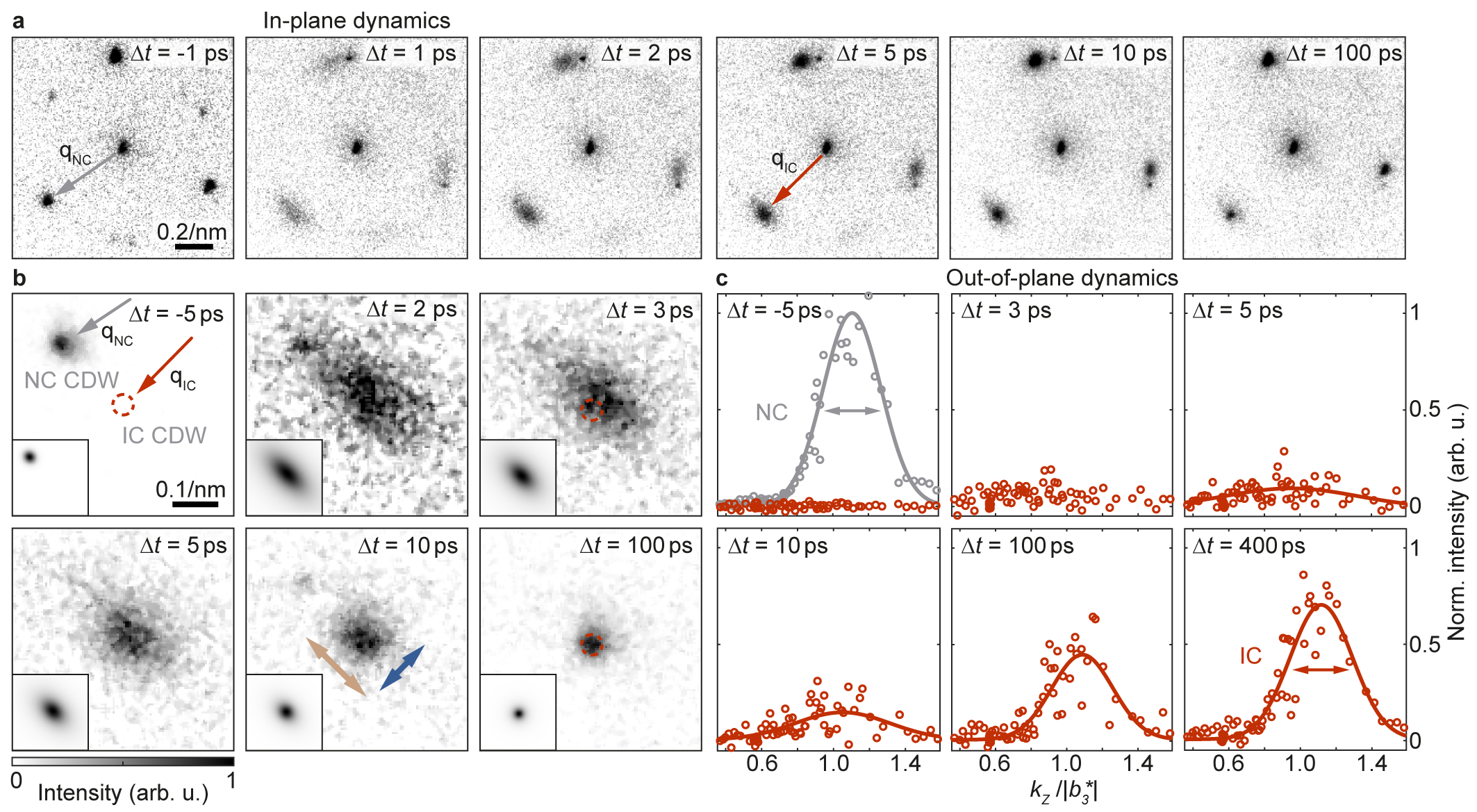}
\caption{
\textbf{Temporal evolution of the in-plane and out-of-plane CDW spot shapes.}
\textbf{a}~Section of reciprocal space in the FOLZ at representative temporal delays $\Delta t$. A triplet of first-order CDW diffraction spots surrounds the associated structural reflection, satellites with lower intensities are higher-order spots of the NC phase attributed to neighbouring reflections.
\textbf{b}~Average over all visible first-order CDW diffraction spots close to the FOLZ. The equilibrium IC spot position is given by dashed red circles. Brown and blue arrows indicate the evaluation axes for the azimuthal and radial spot widths. Insets:~Two-dimensional pseudo-Voigt fit of the spot shape.
\textbf{c}~Out-of-plane spot profiles at representative temporal delays $\Delta t$ reconstructed from in-plane spot intensities measured at different $k_z$ (red circles). The scaled equilibrium spot profile of the NC phase before time-zero is displayed for reference (grey circles). Solid lines represent Gaussian fits of the rocking curves. The spot widths of the NC phase before time-zero and of the IC phase at \SI{400}{\ps} are indicated by arrows and amount to \SI{0.23}{\per\nm} (FWHM), respectively.
}
\label{fig:3}
\end{figure*}

As a consequence, TMDCs susceptible to charge ordering often exhibit more than one thermodynamic CDW phase, governed by a varying balance of, e.g., electron-phonon coupling, electronic correlations and Fermi surface nesting \cite{Rossnagel2011}. These phases may exhibit drastically different macroscopic properties, but only subtle changes of the periodic lattice distortion (PLD) coupled to the CDW \cite{Hellmann2010,Rohwer2011,Yoshida2015}. As demonstrated in a wide range of ultrafast experiments, optical pulses can be used to transiently suppress CDW phases and drive transitions between them \cite{Perfetti2008,Hellmann2010,Rohwer2011,Han2015,Ji2020,Maklar2021,Otto2021,Ravnik2021,Danz2021b,Zhou2021,Eichberger2010,Haupt2016,LeGuyader2017,Vogelgesang2018,Laulhe2017,Zong2018,Jarnac2021,Cheng2022}, induce transient CDW order \cite{Kogar2020} or open up paths into thermodynamically inaccessible hidden states \cite{Stojchevska2014,Han2015,Gerasimenko2019,Stahl2020}. The high degree of in-plane structural disorder induced by the optical excitation not only influences switching behaviour \cite{Vogelgesang2018,Zong2019,Jarnac2021}, but also the final phase texture \cite{Stojchevska2014,Zong2018,Gerasimenko2019}.

The effective dimensionality of a CDW system is determined by the coupling between neighbouring layers. Hence, a phase transition that modifies interlayer correlations may lead to the occurrence of a dimensional crossover \cite{Chen2016,Nicholson2017}. On ultrafast timescales, CDW stacking dynamics have recently been studied by means of ultrafast electron \cite{LeGuyader2017,Cheng2022} and x-ray diffraction \cite{Lantz2017,Jarnac2021}. In particular, an optically induced breakdown of excitonic correlations was shown to quench out-of-plane order \cite{Cheng2022}, and careful tuning of the photoexcitation density was used to realise interlayer phase boundaries with two-dimensional characteristics \cite{Duan2021} -- raising the question if optical control may serve as a gateway to fundamentally low-dimensional phenomena such as the KTHNY transition.

In this article, we report on the observation of transient hexatic order in a nascent incommensurate CDW phase. We use ultrafast high-coherence nanobeam diffraction to track the three-dimensional phase ordering after femtosecond optical excitation. At early times and prior to the establishment of the equilibrium stacking sequence, we identify a quasi-two-dimensional state characterised by a pronounced anisotropic broadening of diffraction spots. We reproduce the experimentally observed behaviour in time-dependent Ginzburg-Landau simulations and attribute it to the presence of unbound topological defects, consistent with the predictions of KTHNY theory.

\section*{Experimental approach and results}

\subsection*{Ultrafast high-coherence nanobeam diffraction}

Over the past two decades, ultrafast electron diffraction (UED) \cite{Siwick2003,Filippetto2022} has evolved into a highly sensitive technique to probe structural phase transitions \cite{Eichberger2010,Haupt2016,LeGuyader2017,Vogelgesang2018,Laulhe2017,Waldecker2017,Zong2018,Sie2019,Kogar2020,Zhou2021,Jarnac2021,Cheng2022} and transient phonon populations \cite{Otto2021,Durr2021,Tauchert2022} on femtosecond time scales. State-of-the-art time-resolved electron diffraction instrumentation typically uses large-diameter beams (\SI{>100}{\um} spot size) for averaged probing of CDW amplitudes, addressing a compromise between a small spot size and sufficient reciprocal-space resolution. Ultrafast Transmission Electron Microscopes (UTEMs) \cite{Zewail2010,Piazza2014,Feist2017,Houdellier2018,Ji2020,Zhu2020a} are versatile devices for ultrafast electron diffraction \cite{VanderVeen2013,Cremons2017}. In particular, laser-triggered field emitters yield UTEM beams of enhanced transverse coherence \cite{Feist2017,Houdellier2018,Zhu2020a}.

As the key experimental innovation of the present study, we harness these capabilities to conduct femtosecond electron diffraction combining high reciprocal-space resolution with a particularly narrow electron beam, in extended series of eight beam tilts. A concomitant transverse coherence length of up to \SI{120}{\nm} enables a precise measurement of in-plane spot profiles and allows to distinguish phases characterised by similar periodicities. Simultaneously, the nanometric probe beam guarantees diffraction from a sample region of sufficient homogeneity in terms of thickness and orientation, as required for the quantitative investigation of stacking dynamics. The fingerprint of such structural modifications is often encoded in low-intensity diffracted signals. We increase the sensitivity to these features by means of a sample design tailored to drive the transformation at an unprecedented laser repetition rate of \SI{610}{\kHz}, maximising the duty cycle of our measurement scheme.

In the experiments (Fig.~\ref{fig:1}), we excite a free-standing \SI{70}{\nm} thin film of the prototypical CDW material \tas{} at room temperature, using ultrashort laser pulses (\SI{800}{\nm} wavelength, \SI{150}{\fs} duration, between \SI{0.3}{\mJ\per\square\cm} and \SI{7.0}{\mJ\per\square\cm} fluence). As a function of a variable temporal delay~$\Delta t$, we capture the transient distribution of CDW spot intensities and profiles using high-coherence ultrashort electron pulses (\SI{120}{\keV} beam energy, \SI{500}{\fs} duration, between \SI{170}{\nm} and \SI{1.1}{\um} spot size, below \SI{0.1}{\milli\radian} convergence semi-angle, see Supplementary Information and Fig.~S3 for fluence-dependent delay scans).

\begin{figure*}[t]
\centering
\includegraphics[width=\textwidth]{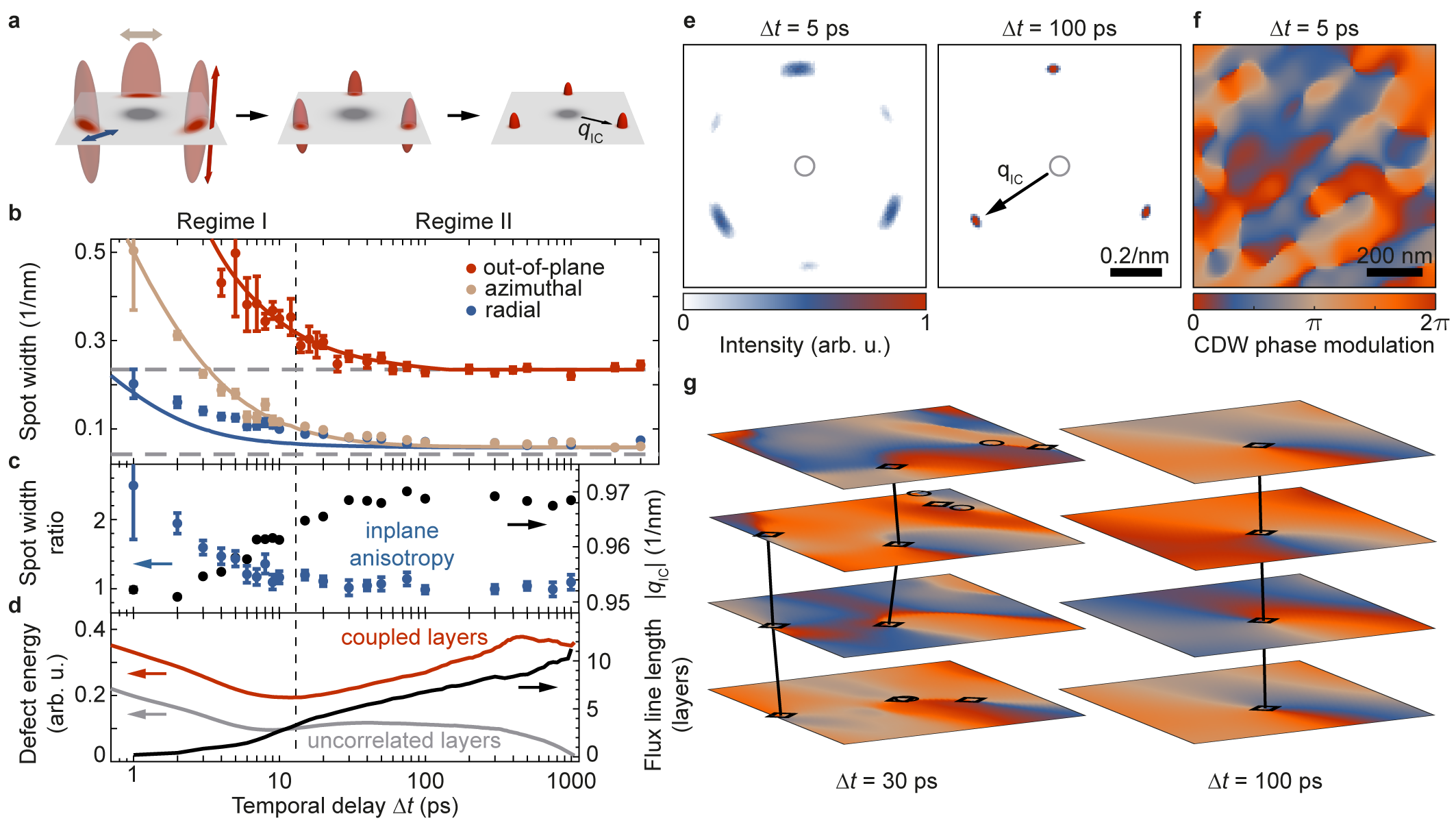}
\caption{
\textbf{Dynamics of the IC spot profile and simulation results.}
\textbf{a}~Schematic temporal evolution of the reciprocal lattice rod shape.
\textbf{b}~Azimuthal (brown) and radial (blue) IC spot width (FWHM) extracted from the in-plane data shown in Fig.~\ref{fig:3}b (see brown and blue arrows at \SI{10}{\ps} delay) and out-of-plane IC spot widths (red) derived from the reconstructed rocking curves shown in Fig.~\ref{fig:3}c. All curves approach the NC equilibrium spot widths at late delay times (dashed grey lines). Solid lines represent the corresponding temporal evolution of the spot widths in the simulation corrected for the instrument resolution (see Methods).
\textbf{c}~Left axis: Temporal evolution of the ratio between azimuthal and radial spot width (blue). Right axis: Dynamics of the CDW wave number $\left|\vec q_{\tiny{\text{IC}}}\right|$ extracted from the fits in Fig.~\ref{fig:3}b (black).
\textbf{d}~Left axis: Simulated energy per phase vortex for a stack of correlated (red) and uncorrelated (grey) layers, respectively. Right axis: Average flux line length corrected for the influence of coincidental alignment present at early times (see Methods).
\textbf{e}~Exemplary simulated diffractograms in a section of reciprocal space as in Fig.~\ref{fig:3}a. In accordance with the experimental results, we find six azimuthally broadened diffraction spots in the FOLZ at early times, while only three CDW reflections indicative of the three-dimensional CDW are present at late times.
\textbf{f}~Simulated phase modulation of $\phi_{3,l}$ within an individual layer.
\textbf{g}~Three-dimensional CDW phase modulation of $\phi_{3,l}$ in 4 of the 51 simulated layers. The stacking sequence favours a local phase shift of $2\pi/3$ between adjacent layers and the alignment of phase vortices into `flux lines' (marked as black circles and diamonds for opposite chirality, respectively). The images display the same region as in \textbf{f}.
}
\label{fig:4}
\end{figure*}

\subsection*{Incommensurate CDW phases in \tas{}}

We investigate the photoinduced transition between two incommensurate CDW superstructures in \tas{} (grey and red regions in Fig.~\ref{fig:1}b). In thermal equilibrium and at temperatures above \SI{353}{\K}, the in-plane modulation wave vectors $\vec q_{i,\text{\tiny IC}}$ (with $i\in \{1,2\}$) of the CDW/PLD are aligned along the lattice vectors $\vec a_i$ of the hexagonal host \cite{Scruby1975}. Due to the presence of a gap-less long-range phase fluctuation (or \emph{phason}) mode, the CDW in this high-temperature incommensurate phase (in short IC phase) is effectively free-floating and only weakly coupled to neighbouring layers \cite{Overhauser1971}. As such, the IC phase in \tas{} is an ideal host for topological defects and, potentially, hexatic ordering \cite{Dai1991}.

Below the phase transition temperature, the modulation remains incommensurate and the wave vectors $\vec q_{i,\text{\tiny NC}}$ form an angle of \SI{\sim 12}{\degree} with the lattice vectors $\vec a_i$ (nearly commensurate or NC phase). With significant contributions from higher harmonics, the real-space structure exhibits a network of discommensurations separated by domains of commensurate character in which the CDW/PLD is locally locked-in with the crystal structure \cite{Scruby1975,Spijkerman1997}. Despite their different in-plane structures, both phases exhibit a commensurate, three-fold stacking periodicity along the out-of-plane direction ($\vec a_3^* = 3\,\vec a_3$, see Fig.~\ref{fig:1}b) \cite{Scruby1975,Spijkerman1997}.

The reciprocal lattice of the host material is given by basis vectors $\vec b_i$. The modulation wave vectors $\vec q_i$ span the reciprocal CDW lattice $(m_1,m_2)$ around every individual main lattice point $(h,k,l)$ (Fig.~\ref{fig:2}a). Hence, each lattice point $\vec k$ can be identified by\cite{Spijkerman1997}
\begin{align}
    \vec k = h\,\vec b_1 + k\,\vec b_2 + l\,\vec b_3^* + m_1\,\vec q_1 + m_2\,\vec q_2.
    \label{eq:Spijkerman}
\end{align}
This notation absorbs the commensurate out-of-plane component of the CDW in a supercell, effectively leading to a shorter reciprocal lattice vector $\vec b_3^* = \vec b_3/3$ and certain systematic absences in diffraction experiments. Specifically, whereas all main reflections (with $m_1,m_2=0$) lie in planes $l=3n$, first-order CDW spots with $(m_1,m_2)=(1,0)$, $(0,1)$ and $(-1,1)$ (and those with opposite sign) are located in planes $l=3n\pm 1$ with non-vanishing $k_z$-components. Thus, they only appear upon tilting the electron beam away from the $[001]$ zone axis \cite{Spijkerman1997,LeGuyader2017}.

A visualisation of the corresponding diffraction geometry is shown in Fig.~\ref{fig:2}b. Under tilted-beam conditions, electron diffractograms typically feature spots in more than one Laue zone (Fig.~\ref{fig:2}d). Main reflections are located in the zero-order Laue zone (ZOLZ) close to the unscattered beam and appear bright compared to the second-order CDW spots surrounding them. As a result of the CDW stacking sequence, first-order CDW reflections are found in the first-order Laue zone (FOLZ) and at larger wave vectors for moderate beam tilts (see also Supplementary Movies S1 and S2).

\subsection*{Dynamics of the in-plane correlation length}

Figure~\ref{fig:3}a shows a series of time-resolved diffraction patterns, averaged over several main reflections in the FOLZ. Before optical excitation (time-zero), we observe a triplet of sharp first-order spots of the initial NC phase. The additional low-intensity reflections in-between stem from higher CDW orders ($>2$). After excitation, the NC spot intensity is largely suppressed within \SI{1}{\ps}. Simultaneously, nearby reflections with $(m_1,m_2)=(1,0)$, $(0,1)$ and $(-1,1)$ emerge, which evidence the nascent IC phase and exhibit a significant anisotropic broadening \cite{Laulhe2017,Vogelgesang2018,Storeck2021}, as well as a few-picosecond increase in intensity \cite{Haupt2016} (cf. also ref. \cite{LeGuyader2017}). Additionally, a slight increase of scattered intensity is detected at $(m_1,m_2)=(-1,0)$, $(0,-1)$ and $(1,-1)$, i.e., opposite of the bright IC spots. Only after these early-stage dynamics, the spot shapes become isotropic, and solely the bright IC spots remain (see image at \SI{100}{\ps}).

The different spot profiles along azimuthal and radial directions (in relation to the nearby main reflection) is a most characteristic hallmark for the presence of a hexatic phase \cite{Nelson1979}. For a quantitative analysis, we recorded a second image series with a longer camera length, i.e., optimised reciprocal-space resolution (Fig.~\ref{fig:3}b). We fit the spot shape at every temporal delay (see insets) and extract the azimuthal and radial spot widths (indicated by brown and blue arrows, respectively). The results are shown in Fig.~\ref{fig:4}b (brown and blue data points). From a spot width ratio $>2$ shortly after optical excitation (blue circles in Fig.~\ref{fig:4}c), the reflections assume an isotropic, yet broadened shape within \SI{10}{\ps} (temporal regime I). On longer timescales, the IC spot width finally approaches that of the NC peak measured before time-zero (temporal regime II; dashed grey line in Fig.~\ref{fig:4}b). Notably, this behaviour coincides with a monotonous growth of the IC wave vector, which initially is \SI{\sim 2}{\percent} shorter than in the equilibrium phase at late delay times (black circles in Fig.~\ref{fig:4}c).

\subsection*{Reconstruction of the CDW rocking curve}

The observed evolution of the in-plane spot profile suggests the presence of hexatic order during temporal regime I. In order to test this hypothesis, we probe the effective dimensionality of the system as induced by the optical excitation, analysing intensities of reflections with different out-of-plane momenta.

Under tilted-beam conditions, the lattice rod is densely sampled across the various spots present in the FOLZ of every single diffractogram (see visualisation in Fig.~\ref{fig:2}b and colour overlay in Fig.~\ref{fig:2}d for the experimental tilt angle of \SI{\sim 2.4}{\degree}). To further enhance sensitivity and fully cover reciprocal space, we perform delay scans with eight different beam tilts. Correlating the scattered intensities before time-zero with dynamical diffraction simulations, we determine the tilt angle for any of these scans with an accuracy better than \SI{0.2}{\degree} (see Supplementary Information and Fig.~S1 for a more detailed description).

Sorting scattered intensities of spots with comparable CDW structure factors (Fig.~\ref{fig:2}c) by their associated out-of-plane components $k_z$, we reconstruct the CDW rocking curve from \SI{11}{} individual IC reflections with a total of \SI{72}{} values of $k_z$ throughout our tilt series at every stage of the dynamics (Fig.~\ref{fig:3}c and Supplementary Information). At early times, the intensity increases independently of $k_z$, indicating a pronounced elongation of the reciprocal lattice rods along the $\vec b_3^*$ direction (see Fig.~\ref{fig:3}c at \SI{3}{\ps} and Supplementary Fig.~S2). Subsequently, a well-defined spot profile emerges and narrows until approaching that of the NC phase measured before time-zero (red data points and dashed grey line in Fig.~\ref{fig:4}b).

\section*{Discussion}

The two CDW modifications involved in the experiment are not commensurate with each other. As a result, there is no continuous global deformation of the crystal during the phase transition. Instead, photoexcitation gives rise to a high density of topological defects, causing the broadening of diffraction spots directly after time-zero. In the past, both CDW dislocations \cite{Vogelgesang2018,Zong2018,Storeck2021,Jarnac2021} and domain walls \cite{Laulhe2017,Jarnac2021} have been invoked to describe the transiently disordered structure on a microscopic scale. While CDW states in \tas{} are known to support both types of topological features \cite{Gerasimenko2019}, the kinetics and interplay between defects in- and out-of-plane remain largely uncertain.

We interpret the evolution of the reconstructed diffraction spots in terms of a transient hexatic phase supported by a temporary suppression of interlayer correlations. Specifically, the observed anisotropic in-plane broadening of CDW reflections (Fig.~\ref{fig:4}d) indicates a pronounced loss of translational order. This feature is characteristic of KTHNY behaviour, relating the condensation of a two-dimensional crystalline phase via a hexatic intermediate to the pair-wise annihilation of point-like defects \cite{Nelson1979}. In our measurements, the occurrence of such a mechanism is further substantiated by the temporal evolution of the IC wave vector (black circles in Fig.~\ref{fig:4}c) whose initial shortening has recently been attributed to a dislocation-induced self-doping of the electronic system \cite{Jarnac2021}. The concurrent, highly uncorrelated out-of-plane structure is apparent from the elongation of reciprocal lattice rods (Fig.~\ref{fig:3}c and Fig.~\ref{fig:4}b), giving rise to the simultaneous visibility of all six IC reflections in the same Laue zone (Fig.~\ref{fig:3}a, \SI{1}{\ps} to \SI{5}{\ps}).

To further explore the notion that a gas of dislocation-type topological defects governs the initial stages of the IC phase formation kinetics, and to understand how three-dimensional ordering affects this process, we implement a multilayer time-dependent Ginzburg-Landau simulation. The underlying free-energy functional is derived from \textsc{Nakanishi's} model \cite{Nakanishi1984} describing the most important CDW phases in \tas{} and their out-of-plane configurations (see Methods for a more detailed description). We express the IC modulation in terms of three coupled order parameters $\phi_{j,l}$ corresponding to the directions of the three in-plane modulation wave vectors $\vec q_{i,\text{\tiny IC}}$ and the layer $l$ in the atomic structure. The physical charge-density modulation (see Supplementary Fig.~S4) is then given by
\begin{align}
    \rho_l(\vec r) = \text{Re}\left[\sum_j\text{exp}\left(2\pi i\,\vec q_j\cdot\vec r\right)\phi_{j,l}\right].
    \label{eq:OP}
\end{align}
This model is capable of closely reproducing the experimentally observed temporal evolution of the reciprocal lattice rod (solid lines in Fig.~\ref{fig:4}b). For the associated set of parameters, the early-stage dynamics are characterised by a high density of uncorrelated phase singularities (Fig.~\ref{fig:4}f). In this state, local energy minimisation leads to a spatially anisotropic character of the phases of the individual order parameters (Fig.~\ref{fig:4}e), causing the typical anisotropic in-plane spot shape shown in Fig.~\ref{fig:3}b. From here on, a pairing of singularities with opposite chirality in the phases of two of the three order parameters leads to the formation of physical CDW dislocations, and enables a local build-up of the PLD amplitude. The evolution of the average free energy associated with an individual phase vortex parallels that of a purely two-dimensional system, as seen for a reference simulation without interlayer coupling (red and grey curves in Fig.~\ref{fig:4}d, respectively). We therefore believe that the observed loss of stacking order at early times is a prerequisite for inducing the intrinsically two-dimensional hexatic intermediate in \tas{}. It is intriguing to speculate whether such a state could in fact be realised as a thermodynamically stable phase during the metallic-to-IC transition in a \tas{} mono-layer.

Beyond temporal delays of \SI{10}{\ps} and after the spot shape anisotropy has decayed, we observe the onset of three-dimensional behaviour. In contrast to the two-dimensional reference system, the energy of individual vortices increases, and their resulting correlation in adjacent layers leads to the formation of `flux lines' (black lines in Fig.~\ref{fig:4}d and g, see Methods for a more detailed description). In parallel, the establishment of long-range order is driven by the annihilation of singularities within each of the $\phi_{j,l}$, eventually reaching correlation lengths at late times that exceed the transverse electron beam coherence length in the experiment.

In conclusion, the identification of hexatic order and stacking dynamics in our study is enabled by ultrafast diffraction with a nanoprobe of exceptional collimation. These results tie in with recent observations of laser-induced transient phases \cite{Ravnik2021} and dimensional cross-overs \cite{Cheng2022,Duan2021}, exemplifying how optical control over interlayer correlations can be used to host low-dimensional states and transitions. Addressing a recurring experimental challenge in ultrafast and materials science, this approach promises to advance high-resolution non-equilibrium investigations in systems characterised by weak structural signatures, sub-micron sample sizes and considerable spatial heterogeneity \cite{Jin2018}. As such, nanoscale structural analysis will guide the design of applications harvesting laser-induced functionality by material composition and tailored responses to external stimuli.

\bibliography{TaS2}

\section*{Methods}

\subsection*{Ultrafast nanobeam diffraction experiments}

The Göttingen UTEM is based on a JEOL JEM-2100F transmission electron microscope modified to enable the investigation of ultrafast dynamics. Femtosecond laser pulses (\SI{515}{\nm} wavelength after frequency doubling of the output of a \enquote{Light Conversion PHAROS} femtosecond laser, \SI{610}{\kHz} repetition rate) are used to generate ultrashort electron pulses from the microscope's ZrO/W Schottky emitter via linear photoemission. A fraction of the laser output is converted to \SI{800}{\nm} wavelength by optical parametric amplification (\enquote{Light Conversion ORPHEUS-F}) and incident on the sample at a variable temporal delay with respect to the electron pulses. Further technical details on the instrumentation are given in ref.~~\cite{Feist2017}.

Snapshots of the non-equilibrium dynamics are recorded on a direct electron detection camera (\enquote{Direct Electron DE-16}) and processed by an electron counting algorithm. The diffractograms presented in this article have been integrated for \SI{3}{\minute} (Fig.~\ref{fig:3}a), \SI{7.5}{\minute} (Fig.~\ref{fig:3}b), and \SI{5}{\minute} (Fig.~\ref{fig:2}d and Fig.~\ref{fig:3}c) per temporal delay, respectively.

\subsection*{Specimen preparation}

The investigated $70\,\text{nm}$ thin film of \tas{} has been obtained by ultramicrotomy. Details on the preparation process and a comprehensive characterisation of the specimen can be found in the Supplementary Information of ref.~~\cite{Danz2021b}.

\subsection*{Time-dependent Ginzburg-Landau simulations}

In the time-dependent Ginzburg-Landau simulations discussed in the main text, we describe the physical CDW modulation $\rho_l$ in layer $l$ of the material as given in equation \ref{eq:OP}.
%
%
This notation effectively strips the equilibrium in-plane CDW periodicity $\vec q_j$ from the additional out-of-equilibrium modulation given by the three coupled order parameters $\phi_{i,l}$. Numerically integrating the equation of motion for the phenomenological free energy $F$ of the system
\begin{align}
    \frac{\delta \phi_{j,l}}{\delta t}=-\Gamma\frac{\delta F}{\delta \phi_{j,l}^*}
\end{align}
then yields the spatiotemporal dynamics shown in Fig.~\ref{fig:4}, starting from a stack of uncorrelated layers where both amplitude and phase of the $\phi_{j,l}$ are randomised. The parameter $\Gamma$ controls the overall timescale of the dynamics. For the free-energy functional, we choose an approach tailored to model the phase diagram of the different three-dimensional CDW configurations in \tas{} \cite{Nakanishi1984}, i.e.,
\begin{align}
    \begin{split}
    F &=\sum_l\int\text{d}^2r\bigg[\sum_j \bigg(\phi_{j,l}^*A_j(\vec q_j-i \nabla)\phi_{j,l}+B\left|\phi_{j,l}\right|^4\\
    & +C\left|\phi_{j,l}\phi_{j+1,l}\right|^2+E\,\text{Re}(\phi_{j,l}^3\phi_{j+1,l}^*)\bigg)\\
    & +D\,\text{Re}(\phi_{1,l} \phi_{2,l} \phi_{3,l})\bigg]\\
    & +\sum_l G \int\text{d}^2r\sum_j\,\text{Re}\big[e^{ig_1}\phi_{j,l}^*\phi_{j,l+1}\\
    & +ae^{ig_2}\phi_{j,l}^*\phi_{j,l+2}\big].
    \end{split}
    \label{eq:free-energy}
\end{align}
Therein, energy minimisation of the $B$ and the $C$-term ensures a local equilibration of the CDW amplitude and the phasing-term $D$ gives the relative phase relation between the order parameters such that well-defined CDW maxima emerge in a hexagonal arrangement (see Supplementary Fig.~S4). The transition from a local lock-in of the CDW with the underlying main lattice in the C-phase to the incommensurate modulations found above a critical temperature $T_C$ is governed by a temperature-dependent competition between the commensurability energy $E$ and the kinetic energy $A$. In an out-of-equilibrium scenario, a minimisation of the latter also determines the kinetics of topological defects. In order to account for the temperature-dependent relative orientation between the NC CDW wave vector and the main lattice periodicities \cite{Scruby1975}, the kinetic energy includes a softness towards a distortion of the order parameter along the azimuthal component, which is in conceptual agreement with the modelling of the KTHNY transition of a two-dimensional solid \cite{Nelson1979}. One finds \cite{Nakanishi1984}:
\begin{align}
    A_j\left(\vec q_j,T\right) = T-T_C+(1-\xi_j)u+\xi_j v(1-\text{cos}\,6\varphi_j)
\end{align}
with
\begin{align}
    \xi_j=1-s(|\vec q|-|\vec q_j|)^2/|\vec b_j|^2
\end{align}
where $\varphi_j$ describes the angle between the wave vector $\vec q$ and $\vec q_j$, $\vec b_j$ are the reciprocal lattice vectors of the undistorted structure and $u$, $v$ and $s$ are parameters.

Along the out-of-plane components, equation \ref{eq:free-energy} perturbationally treats the coupling of the individual layers to their nearest neighbouring and next-nearest neighbouring layer via the parameters $G$ and $a$, respectively, while the finite phase factors $g_1$ and $g_2$ ensure the establishment of the expected stacking periodicity \cite{Nakanishi1984}.

Starting from a randomised phase and amplitude pattern in all $\phi_{j,l}$, we find that a parameter set with $B=\SI{5e-4}{}$, $C=\SI{1e-3}{}$, $D=\SI{-5e-4}{}$, $E=\SI{-7.5e-6}{}$, $s=\SI{6}{}$, $u=\SI{1.3}{}$, $v=\SI{0.4}{}$, $G=\SI{0.4}{}$, $a=\SI{0.5}{}$, $g_1=\SI{-0.7}{}$, $g_2=\SI{0.7}{}$ and $\Gamma=\SI{1.43e-2}{}$ reproduces our experimental results. Our simulation volume consists of \SI{51}{} individual layers with approximately $\SI{250}{}\times\SI{290}{}$ CDW unit cells along the lateral dimensions each.

In order to extract spot widths from the simulation, we perform a three-dimensional Fourier-transformation of the real parts of the $\phi_{j}$ at representative stages of the temporal dynamics to derive the simulated reciprocal lattice rod. Summing the squared modulus of the momentum distribution in planes corresponding to the out-of-plane momentum $|\vec b_3^*|$ (considering the different rotations of the respective $\vec q_j$) for all three $\phi_{j}$ then gives the simulation analogue to the experimental diffraction spots depicted in Fig.~\ref{fig:3}b. Fitting a two-dimensional Lorentzian function to this in-plane spot-profile yields the corresponding spot widths $\gamma_{\text{sim}}$ along the azimuthal and radial directions. Along the out-of-plane directions, we assign the integrated intensity of the CDW diffraction spot in every reciprocal lattice plane to its out-of-plane component and fit a Gaussian function centred at the corresponding equilibrium stacking periodicities $|\vec b_3^*|$ to the resulting one-dimensional intensity distribution. The derived spot widths $\gamma$ (FWHM) over the course of the dynamics are displayed in Fig.~\ref{fig:4}b. For adequate comparison between simulation and experimental results, we include the instrument resolution $\gamma_r$ in the form $\gamma=\sqrt{\gamma_{\text{sim}}^2+\gamma_r.^2}$ where $\gamma_r$ is the corresponding spot width measured at late times.

Within this model, the phase formation is governed by the kinetics of point-like defects that appear as vortices in the $\phi_{j,l}$. For the `flux lines' length shown in Fig.~\ref{fig:4}d, we assume vortices in neighbouring layers as correlated when their spatial separation along the in-plane coordinates is less than five CDW lattice vectors. At early times, this definition of the interlayer vortex correlation is additionally influenced by coincidental alignment stemming from the high initial density of defects within every layer. We correct for this influence by subtracting the temporal evolution of the `flux line' length derived in the same manner for the case of uncoupled CDW layers.

The average energy of an individual vortex for both the coupled and uncorrelated stack of CDW layers is derived by integrating equation \ref{eq:free-energy} at every step of the dynamics and additionally considering the ground-state energy of the equivalent fully equilibrated system. For the latter, we chose a uniform amplitude within the $\phi_{j,l}$ and a relative phase shift of $2\pi/3$ between neighbouring layers as the initial condition and let the local amplitude relax until the system reaches its energy minimum.

\section*{Acknowledgements}

The authors thank M. Sivis for technical support in focused ion beam milling in the specimen preparation and C. Wichmann for providing the specimen holder. Furthermore, we gratefully acknowledge useful discussions with A.~Zippelius as well as support from the Göttingen UTEM team, especially J.H. Gaida, M. Möller and K. Ahlborn.
This work was funded by the Deutsche Forschungsgemeinschaft (DFG, German Research Foundation) in the Collaborative Research Centre ``Atomic scale control of energy conversion'' (217133147/SFB 1073, project A05) and via resources from the Gottfried Wilhelm Leibniz Prize (RO 3936/4-1). Th.D. gratefully acknowledges a scholarship by the German Academic Scholarship Foundation.

\section*{Author contributions}

Ti.D. and Th.D. conducted the experiments, analysed the data, and prepared the specimen. Ti.D. conducted the diffraction simulations and the fitting of the specimen tilt angle with support from S.S. S.Y. wrote the Ginzburg-Landau simulations with input from Ti.D. and Th.D. K.R. provided high-quality \tas{} crystals. C.R. conceived and directed the study. All authors discussed the results and their interpretation. Ti.D., Th.D. and C.R. wrote the manuscript with discussions and input from all authors.

\section*{Competing interests}

The authors declare no competing interests.

\end{document}


\title{\sffamily\Large\texorpdfstring
{Light-induced hexatic state in a layered quantum material}
{Light-induced hexatic state in a layered quantum material}
}

\author{Till Domröse}
\author{Thomas Danz}
\affiliation{Max Planck Institute for Multidisciplinary Sciences, 37077 Göttingen, Germany}
\author{Sophie F. Schaible}
\affiliation{4th Physical Institute -- Solids and Nanostructures, University of Göttingen, 37077 Göttingen, Germany}
\author{Kai Rossnagel}
\affiliation{Institute of Experimental and Applied Physics, Kiel University, 24098 Kiel, Germany}
\affiliation{Ruprecht Haensel Laboratory, Deutsches Elektronen-Synchrotron DESY, 22607 Hamburg, Germany}
\author{Sergey V. Yalunin}
\affiliation{Max Planck Institute for Multidisciplinary Sciences, 37077 Göttingen, Germany}
\author{Claus Ropers}
\email[Corresponding author: ]{claus.ropers@mpinat.mpg.de}
\affiliation{Max Planck Institute for Multidisciplinary Sciences, 37077 Göttingen, Germany}
\affiliation{4th Physical Institute -- Solids and Nanostructures, University of Göttingen, 37077 Göttingen, Germany}

\maketitle

\section*{Reconstruction and fitting of CDW rocking curves}

\begin{figure*}[t]
\centering
\includegraphics[width=\textwidth]{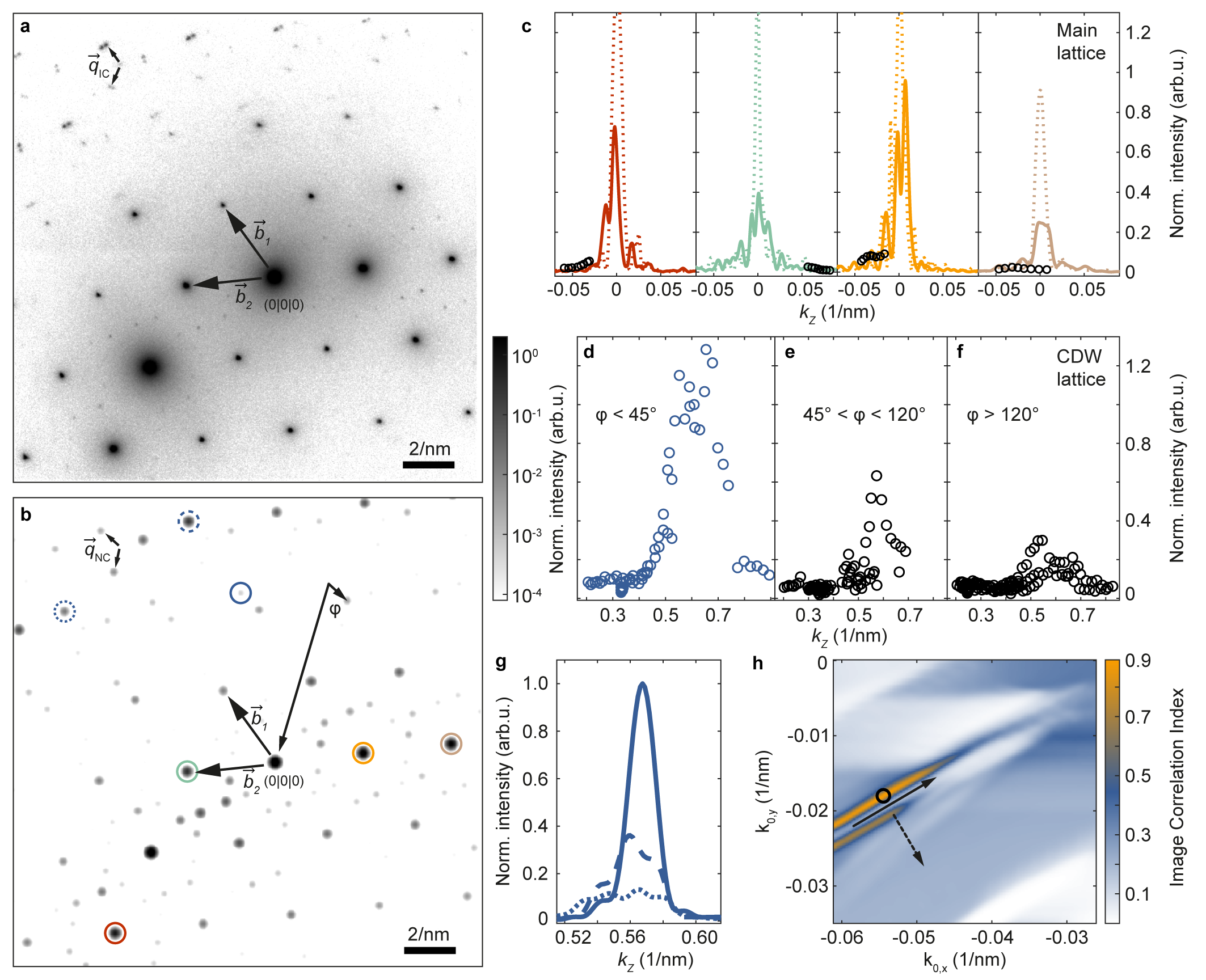}
\caption{
\textbf{Fitting of specimen tilt angle and reconstructed rocking curves.}
\textbf{a}~Exemplary electron diffractogram recorded with one of the overall eight different specimen tilt angles used to reconstruct CDW rocking curves. The displayed image is a temporal average over all stages of the phase formation kinetics as can be recognised from the presence of spots corresponding to both CDW phases. Direct electron detection enables simultaneous recording of scattered intensities spanning five orders of magnitude (see intensity scale).
\textbf{b}~Simulated electron diffractogram for the NC phase of \tas{} for the fitted tilt angle in \textbf{a}. Coloured circles mark the spots belonging to the measured and simulated rocking curves in \textbf{c}, \textbf{d} and \textbf{g}, respectively. Scattered CDW intensity is additionally modulated by the CDW structure factor which predominantly depends on the angle $\varphi$ (cf. equation \ref{eq:Overhauser}).
\textbf{c}~Measured (black circles) and simulated (solid lines) rocking curves of the main lattice reflections indicated in \textbf{b}. The dotted curves describe the rocking curve for an additional sample tilt of \SI{0.29}{\degree} along the direction indicated by the dotted arrow in \textbf{h}, exemplifying the pronounced influence of dynamical scattering on bright intensity reflections situated in the ZOLZ. Measured data is rescaled by a global factor for all curves to match the simulated intensities.
\textbf{d}-\textbf{f} Measured NC CDW rocking curves for the fitted specimen tilt angle and sorted by the CDW structure factor (cf. equation \ref{eq:Overhauser}). The bright-intensity CDW spots in \textbf{d} are used to extract the out-of-plane shape of the reciprocal lattice rod discussed in the main text. Especially CDW spots with smaller structure factor (\textbf{e} and \textbf{f}) are subject to dynamical scattering.
\textbf{g}~Simulated CDW rocking curves for the spots highlighted in \textbf{b}. Overall, dynamical scattering  has a smaller influence on first order CDW intensities than on main lattice reflections in the ZOLZ. Still, it causes an additional broadening of the spot profile and a shift of the $k_z$-component with highest CDW intensity (dotted curves).
\textbf{h}~Fitted quality factors of the template matching described in the text as a function of the $k_x,k_y$-components of the normalised incident wave vector $k_0$. The black circle indicates the fitting results, the tilt direction for the rocking curves in \textbf{c} and \textbf{g} is given by the solid black arrow. A similar virtual tilt series only shifted by \SI{0.29}{\degree} along the direction given by the dotted arrow yields the dotted rocking curves displayed in \textbf{c}.
}
\label{fig:S1}
\end{figure*}

In our experiments, we reconstruct the out-of-plane shape of the CDW reciprocal lattice rod by assigning the respective $k_z$-component to the measured intensity of every CDW diffraction spot in each individual image of the image series. In addition to a modulation by $k_z$, the intensity of a CDW spot of order $n$ at a scattering vector $\vec k$ depends on the structure factor $V_{\vec{k}}$, which can be expressed in terms of a Bessel function of first kind $J_n(x)$: \cite{Overhauser1971}

\begin{align}
    V(\vec{k}) \propto J_n(2\pi\,\vec{k}\cdot\vec{A}).
    \label{eq:Overhauser}
\end{align}
For the longitudinal distortion with comparably small amplitude $\vec{A_i}$ found here ($\vec A_i\,||\,\vec q_i$, where $\vec q_i$ is the CDW wave vector)\cite{Scruby1975} and considering high-energy electrons, the diffractograms mainly contain information about the modulation of the tantalum sublattice \cite{Colliex2006}. Additionally, CDW spots, to a good approximation, can be sorted by their respective scattered intensity into three groups such that the CDW rocking curve can be reconstructed from multiple CDW spots within a single image (Fig.~\ref{fig:S1}d-f).

\begin{figure*}[t]
\centering
\includegraphics[width=\textwidth]{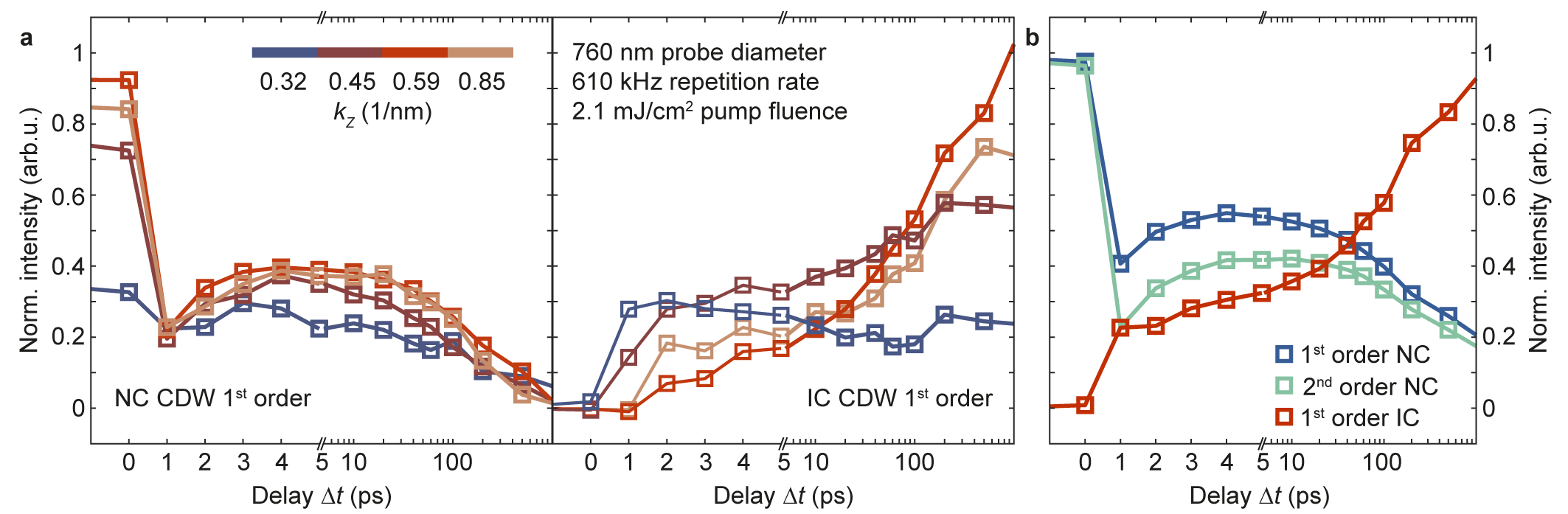}
\caption{
\textbf{CDW diffracted intensity for reconstruction of rocking curves.}
\textbf{a}~Intensity of first-order NC (left) and IC (right) diffraction spots as a function of temporal delay for the data set shown in Fig.~3 in the main text. The curves are additionally sorted and binned by the $k_z$-component of the respective reflections. The scattered intensity at the IC spot position before time-zero is subtracted for both types of curves as a measure of the inelastic background in the images. The NC curves are additionally normalised with respect to the measured intensity before time-zero and show similar behaviour as a function of temporal delay (cf. Fig.~3b). For the IC curves, we normalise to the measured intensity after \SI{300}{\ps}. During the phase formation, IC spots probed at a $k_z$-component closer to the ZOLZ display an initial overshoot (blue curve). In contrast, reflections in the FOLZ are characterised by a continuous intensity increase (red curve), leading to the sharpening of the reciprocal lattice rod along the out-of-plane direction described in the main text and Fig.~3.
\textbf{b}~Average intensity of first-order NC (blue), first-order IC (red) and second-order NC spots (green). Curve normalisation as described in \textbf{a}.
}
\label{fig:S2}
\end{figure*}

For the CDW rocking curves displayed in the main text, we only consider the brightest reflections where the angle $\varphi$ between $\vec k$ and $\vec q_i$ (see Fig.\ref{fig:S1}b) is smaller than \SI{45}{\degree} (see Fig.~\ref{fig:S1}d for the resulting NC rocking curve at $\Delta t<\SI{0}{\ps}$). In comparison, CDW reflections with $\varphi>\SI{45}{\degree}$ are less intense and more heavily affected by dynamical scattering effects (Fig.~\ref{fig:S1}e-f). The reconstruction of the CDW rocking curves is based on the $k_z$-dependent temporal evolution of scattered CDW intensities shown in Fig.~\ref{fig:S2}. Following optical excitation, the NC CDW amplitude is suppressed within \SI{1}{\ps} irrespective of the out-of-plane component. In contrast, IC diffraction spots situated at $k_z$-components deviating from the FOLZ display an initial intensity overshoot, whereas those spots probed directly in the FOLZ experience a monotonous increase of scattered intensity over the entire temporal regime (Fig.~\ref{fig:S2}a).

The length of the reciprocal lattice rod at every temporal delay is determined by fitting a Gaussian-shaped profile to the rocking curves inferred from these delay curves. Every rocking curve consists of intensities scattered into \SI{11}{} individual IC reflections probed at varying $k_z$ in our tilt series, resulting in a total of \SI{72}{} out-of-plane momenta considered for the fitting of the rod shape. The majority of these CDW reflections lie at out-of-plane wave vector components smaller than the equilibrium stacking periodicity of the CDW. In order to enhance the fitting weight of the brightest spots close to the centre of the rocking curve, we linearly interpolate the measured CDW intensity to gain a homogeneously spaced distribution along $k_z$. The fit results with the corresponding error bars are displayed in Fig.~3c and Fig.~4b in the main text.

\section*{Dynamical electron diffraction simulations}

Within a single diffractogram, scattered CDW intensities depend on the excitation error, i.e., the mismatch to an exactly fulfilled Laue condition, that is determined by the relative tilt angle between specimen and electron beam. In order to determine this tilt angle from measured intensities in our tilt-series, we perform dynamical diffraction simulations based on the scattering matrix approach. Therein, the incident electron wave function inside the crystal is divided into individual beams $n$ scattered by a respective crystal periodicity $\vec k_n$ \cite{DeGraef2003}. The beam amplitude $\Psi_n$ as a function of the penetration depth $z$ can then be described as
\begin{align}
    \frac{\text{d}\Psi_n}{\text{d}z} = i\pi\sum_{n'=1}^{N}\left[2\,s_{n'}\,\delta\left(n-n'\right)+\xi_{n,n'}\right]\Psi_{n'}\, .
    \label{eq:DHW}
\end{align}
Equation \ref{eq:DHW} is the \textit{Darwin-Howie-Whelan equation} for elastic scattering of high-energy electrons. It can be expressed in terms of a scattering matrix $S_N$ \cite{Fujimoto1959} and then integrated numerically in close analogy to the treatment of $N$ coupled harmonic oscillators. The diagonal elements of this matrix account for the scattering geometry in the form of the excitation error $s_n$, describing the deviation from an exactly fulfilled Laue condition. The transfer of scattering amplitude between beams $n$ and $n'$ is mediated by the off-diagonal elements $\xi_{n,n'}$ representing the effective scattering length, thereby incorporating the influence of the specimen. Specifically, $\frac{1}{\xi_{\vec{k}}}\propto\lambda\lvert V_{\vec{k}} \rvert$ where $\lambda$ is the electron wavelength and $V_{\vec{k}}$ the Fourier component of the scattering crystal potential, i.e., the structure factor associated with the wave vector $\vec{k}$.

Our simulations are based on the reported crystal structure of the NC phase \cite{Spijkerman1997} where we approximate the incommensurate lattice distortion by a supercell comprised of $45\times 45\times 3$ unit cells of the undistorted structure (see ref. \cite{danzUltrafastTransmissionElectron2021} for a justification of this approximation), while the out-of-plane periodicity of the NC phase is given by the $\langle\sigma_{h2}\rangle$ stacking described in detail in ref. \cite{Nakanishi1984}. The structure factor of the material is calculated with the atomic scattering factors for tantalum and sulphur listed in table~\ref{tab:param} \cite{Colliex2006}, yielding scattering lengths of, e.g., $\xi=\SI{108}{\nm}$ for a scattering vector relating a first-order main lattice reflection to the direct beam, and $\xi=\SI{1545}{\nm}$ for a first-order NC spot relative to its associated main lattice spot.

In our simulations, we consider $61$ main lattice reflections situated in the ZOLZ, in addition to a total of $786$ CDW spots of first and second-order in the FOLZ and in the ZOLZ, respectively. Thereof, only those wave vectors yielding a scattering strength of more than \SI{2}{\percent} of the maximum found at $\vec q=0$ are taken into account. Furthermore, we include the effects of finite temperatures on scattered intensities in the form of the reported Debye-Waller factors of tantalum and sulphur at room-temperature \cite{BalaguruRayappan2010}. A numerical integration of equation \ref{eq:DHW} with varying sample thickness and electron beam incidence then allows us to fit the overall specimen tilt following the approach described in the subsequent section.

A resulting simulated diffractogram for an exemplary beam incidence from our measurements is depicted in Fig.~\ref{fig:S1}b. We account for a spatially varying sample orientation and the finite electron beam convergence angle by averaging over beam tilts within a range  of \SI{1.5}{\milli\radian} centred around the fitted beam tilt. Since electron diffractograms, especially those recorded close to zone-axis orientation, always feature multiple diffraction spots of significant intensity, the average \SI{70}{\nm} thickness of our sample results in non-negligible effects of dynamical scattering on the measured intensities, leading to pronounced deviations from kinematical scattering theory. In particular, diffraction spots of the undistorted lattice situated in the vicinity of the Laue circle in the ZOLZ display strong intensity oscillations during a (virtual) beam tilt series, and, additionally, are highly susceptible to small variations of the tilt direction (dotted rocking curves in Fig.~\ref{fig:S1}c). In contrast, first-order CDW reflections found at larger wave vectors are less affected by multiple scattering events (Fig.~\ref{fig:S1}g). For these spots, the simulations yield only minor intensity modulations along $k_z$ and a small shift of the rod centre.

\begin{table}[b]
\centering
\begin{tabularx}{\columnwidth}
{ 
  | >{\centering\arraybackslash\hsize=.5\hsize}X 
  | >{\centering\arraybackslash\hsize=.25\hsize}X
  | >{\centering\arraybackslash\hsize=1.625\hsize}X 
  | >{\centering\arraybackslash\hsize=1.625\hsize}X
  | >{\centering\arraybackslash\hsize=1\hsize}X |}
\hline
    & $i$ & $a_i~\left[\SI{}{\angstrom}\right]$\cite{Colliex2006} &  $b_i~[\SI{}{\angstrom^2}]$\cite{Colliex2006} & $W [\SI{}{\angstrom^2}]$\cite{BalaguruRayappan2010}    \\
 \hline
 \hline
   & 1 & \SI{0.3835}{} & \SI{0.0810}{} &   \\
    & 2 & \SI{1.6747}{} & \SI{0.8020}{} &   \\
  Ta & 3 & \SI{3.2986}{} & \SI{4.3545}{} &   0.1\\
    & 4 & \SI{4.0462}{} & \SI{19.9644}{} &   \\
    & 5 & \SI{3.4303}{} & \SI{73.6337}{} &   \\
 \hline
    & 1 & \SI{0.0915}{} & \SI{0.0838}{} &   \\
    & 2 & \SI{0.4312}{} & \SI{0.7788}{} &   \\
  S & 3 & \SI{1.0847}{} & \SI{4.3462}{} &   2\\
    & 4 & \SI{2.4671}{} & \SI{15.5846}{} &   \\
    & 5 & \SI{1.0852}{} & \SI{44.6365}{} &   \\
 \hline
\end{tabularx}
\caption{\textbf{Parameters used in the diffraction simulation.} The atomic scattering factors $f$ of tantalum and sulphur to calculate the structure factor for scattering by a wave vector with magnitude $k$ are derived following $f=\sum_i a_i\,\text{exp}\left(-b_ik^2\right)$. The coefficients apply to $k<\SI{6}{\per\angstrom}$. The structure factor is additionally modulated by the Debye-Waller factor $\text{exp}\left(-2Wk^2\right)$, the specified values for $W$ apply to a broad temperature range around $\approx\SI{300}{\kelvin}$.}
\label{tab:param}
\end{table}

\section*{Fitting of specimen tilt angles}

In any TEM, the electron beam angle of incidence can be set by beam deflectors, the resulting shift of the diffractogram yields a reliable measure of the relative beam tilt between two settings. The absolute beam tilt, however, is encoded in diffracted intensities and requires fitting routines based on diffraction simulations \cite{Rauch2014}.

Using the simulation approach outlined in the previous section, we calculated more than \SI{32000}{} diffractograms with variable angles of incidence, while averaging over the thickness distribution of our specimen (see supplementary material of ref. \cite{Danz2021b}) and a relative angle of incidence between specimen and electron beam \SI{1.5}{\milli\radian}, considering both the electron beam convergence angle as well as a spatial variations of the sample orientation. The simulated intensity $T_{i,j}$ of a diffraction spot $j$ for every incidence $i$ is then compared to the experimental intensity $P_j$ measured before time-zero.

The strong impact of dynamical scattering leads to ambiguous fitting results for the individual $T_i$. To overcome this limitation, we additionally make use of the known relative beam tilts $\Delta$ in our tilt series, consisting of delay scans recorded under eight different angles of incidence. Summing over the intensities of all main lattice and first-order CDW reflections, this approach yields the image correlation index \cite{Rauch2014}
\begin{align}
 Q_i = \prod_{\Delta=1}^8\frac{\sum_j^m P_j\left(\Delta\right)\,T_{i,j}\left(\Delta\right)}{\sqrt{\sum_j^m P_j^2\left(\Delta\right)}\sqrt{\sum_j^m T_{i,j}^2\left(\Delta\right)}}\, ,
 \label{eq:TemplateMatching}
\end{align}
where $Q_i=1$ indicates a perfect agreement between experiments and simulation.

The calculated $Q_i$ are depicted in Fig.~\ref{fig:S1}h. We derive a maximum correlation index of $Q_i=\SI{0.89}{}$. The corresponding beam tilt results in the simulated intensities of main lattice spots shown in Fig.~\ref{fig:S1}c and yields rocking curves for first-order CDW reflections that are centred around the wave vector of the equilibrium CDW stacking sequence (Fig.~\ref{fig:S1}d-f).

\section*{Fluence-dependent ultrafast nanobeam diffraction}

\begin{figure*}[t]
\centering
\includegraphics[width=\textwidth]{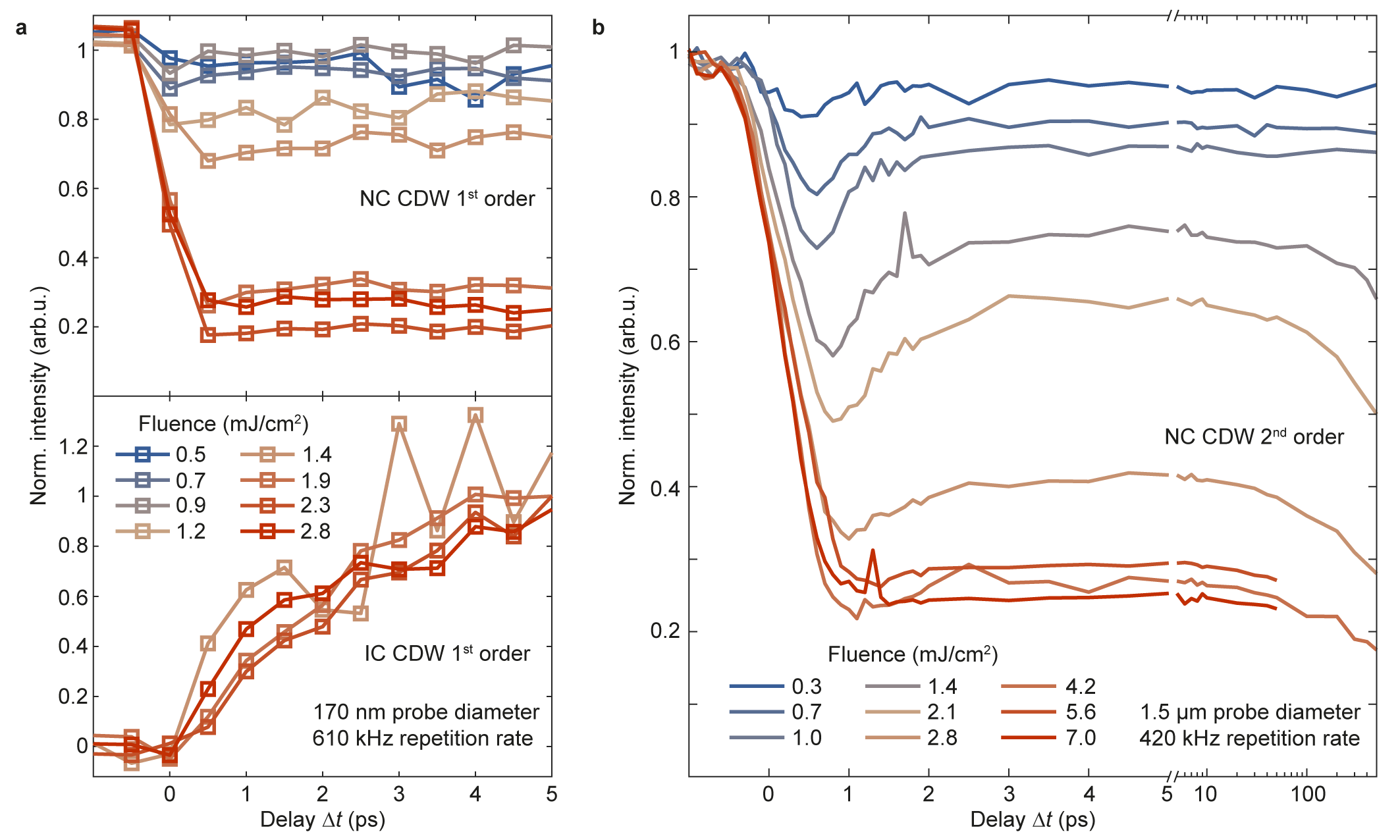}
\caption{
\textbf{Ultrafast nanobeam diffraction of CDW dynamics.}
\textbf{a}~Intensity of first-order NC (top) and IC (bottom) diffraction spots as a function of temporal delay for varying pump fluence. The data was recorded under tilted-sample conditions with an electron beam size of \SI{170}{\nm} and a repetition rate of \SI{610}{\kHz}. The curves are normalised as in Fig.~\ref{fig:S2}. We observe an initial suppression within \SI{500}{\fs} that scales with the pump fluence. Above a threshold of around \SI{2}{\mJ\per\cm^2}, the measured NC intensity after time-zero equals the inelastically scattered background. For such higher fluences, we measure an increase of IC spot intensity within \SI{5}{\ps} with a temporal characteristic independent of the pump fluence.
\textbf{b}~Delay scans of second-order NC intensity for varying pump fluence with the sample tilted into the zone-axis. In contrast to \textbf{a}, the electron beam has a larger diameter (\SI{1.5}{\um}), adding a spatiotemporal component to the dynamics following the initial suppression. For intermediate fluences, we witness a partial recovery of NC intensity after around \SI{5}{\ps}, followed by a second suppression at later delays. We attribute this behaviour to a locally varying excitation density leading to the formation and subsequent growth of IC phase domains on longer timescales as described in ref. \cite{Danz2021b}.
}
\label{fig:S3}
\end{figure*}

In our experiments, we employ a sample design (described in more detail in ref.~\cite{Danz2021b}) that is tailored to minimise the cumulative heating of our \tas{} specimen. This enables laser pump/electron probe experiments at exceptionally high laser repetition rates (up to \SI{610}{\kHz}) over a wide range of pump fluences (here: up to \SI{7.0}{\mJ\per\square\cm}). The thusly enhanced duty cycle in our measurements allows us to further limit the diameter of the collimated electron beam by apertures down to \SI{170}{\nm} in the specimen plane. The resulting delay curves of CDW intensities scattered into first and second-order CDW spots are shown in Fig.~\ref{fig:S3}a and b, respectively. Following the optical excitation, intensity scattered into both first and second-order NC CDW spots is suppressed within \SI{1}{\ps}. For higher pump fluences, IC CDW spots emerge on a similar timescale.

\bibliography{TaS2}

\clearpage
\onecolumngrid
\noindent \textbf{Movie S1}~\textbf{Electron beam tilt-series around the ZOLZ}. Tilting the electron beam around the ZOLZ modulates diffracted intensities. For higher beam tilts, first-order CDW reflections appear at larger wave vectors.
\vspace{5cm}

\noindent \textbf{Movie S2}~\textbf{Electron beam tilt-series around the FOLZ}. Higher beam tilts shift the probing of first-order CDW reflections to smaller wave vectors.

\clearpage

\clearpage

\clearpage

\begin{figure*}[t]
\centering
\includegraphics[width=\textwidth]{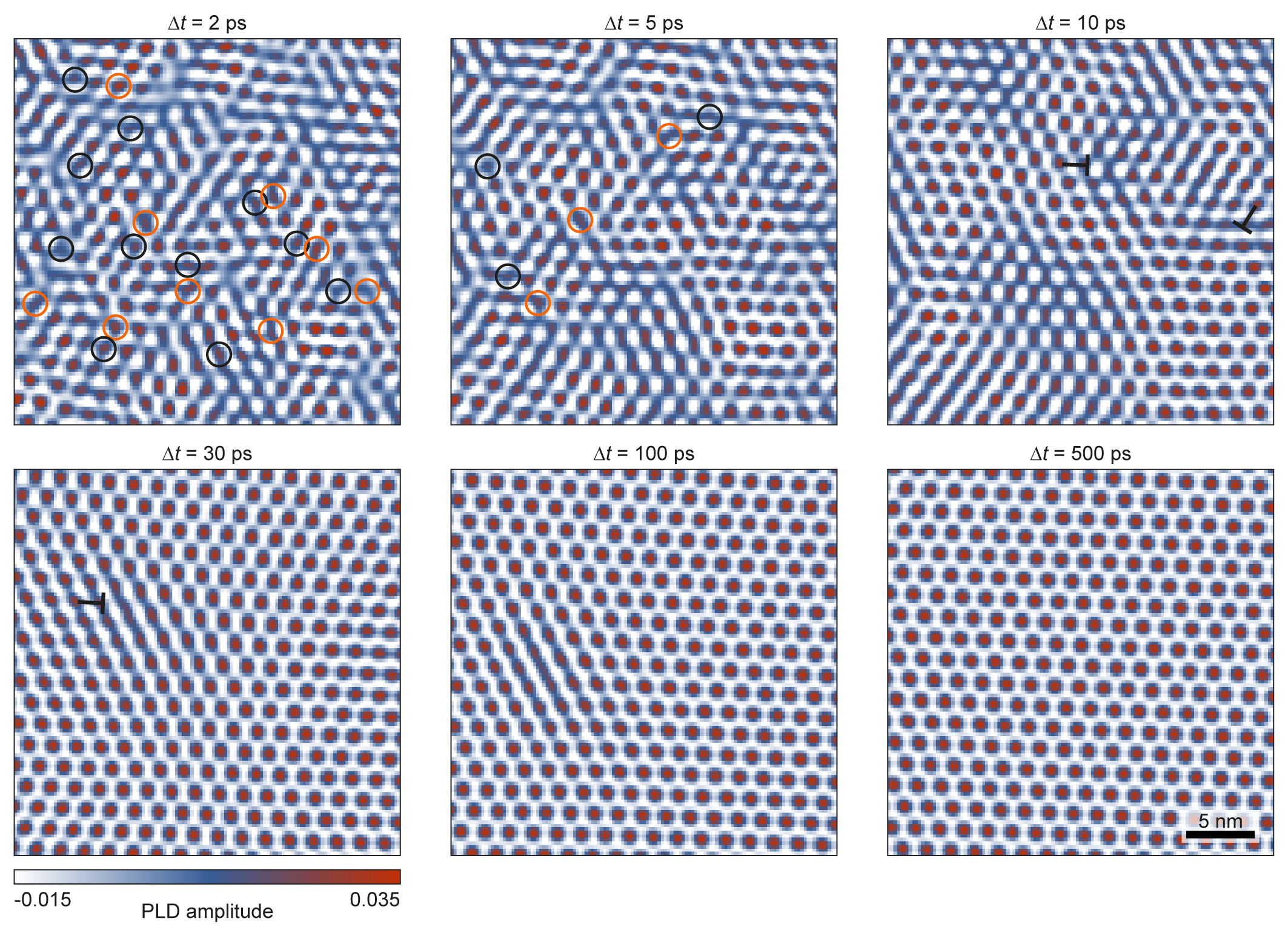}
\caption{
\textbf{CDW formation in time-dependent Ginzburg-Landau simulations.}
CDW pattern $\rho_l$ (see main text) at variable temporal delay derived from the time-dependent Ginzburg-Landau simulations. Orange and black circles highlight several disclinations with seven-fold and five-fold coordination, respectively (\SI{2}{\ps} and \SI{5}{\ps} temporal delay). The late stage dynamics is governed by the annihilation of dislocations (black markers at \SI{10}{\ps} and \SI{30}{\ps} temporal delay).
}
\label{fig:S4}
\end{figure*}